\documentclass[twocolumn,showpacs,preprintnumbers,amsmath,amssymb]{revtex4}

\usepackage{graphicx}
\usepackage{dcolumn}
\usepackage{bm}

\def\bea{\begin{eqnarray}}
\def\eea{\end{eqnarray}}

\def\pp{\mbox{$p$-$p$} }
\def\auau{\mbox{Au-Au} }
\def\cucu{\mbox{Cu-Cu} }
\def\pbpb{\mbox{Pb-Pb} }
\def\aa{\mbox{A-A} }
\def\nn{\mbox{N-N} }

\begin{document} 

\preprint{Version 1.5}

\title{Comparing the same-side ``ridge'' in CMS \pp angular correlations to RHIC \pp data
}

\author{Thomas A. Trainor and David T. Kettler}
\address{CENPA 354290, University of Washington, Seattle, WA 98195}


\date{\today}

\begin{abstract}
The CMS collaboration has recently reported the appearance of a same-side ``ridge'' structure in two-particle angular correlations from 7 TeV \pp collisions. The ridge in \pp collisions at 7 TeV has been compared to a ridge structure in more-central \auau collisions at 0.2 TeV interpreted by some as evidence for  a dense, flowing QCD medium. In this study  we make a detailed comparison between 0.2 TeV \pp correlations and the CMS results. We find that 7 TeV minimum-bias jet correlations are remarkably similar to those at 0.2 TeV, even to the details of the same-side peak geometry. Extrapolation of azimuth quadrupole systematics from 0.2 TeV suggests that the same-side ridge at 7 TeV is a manifestation of the azimuth quadrupole with amplitude enhanced by applied cuts.
\end{abstract}


\maketitle


 \section{Introduction}

The CMS collaboration has recently released the first study of angular correlations from \pp collisions at the Large Hadron Collider (LHC)~\cite{cms}. A notable result is the appearance of a same-side ``ridge'' structure possibly related to the expected jet peak at the angular origin (intrajet correlations). The same-side ridge observed in \pp collisions at 7 TeV has been compared to a ridge structure observed in more-central \auau collisions at the Relativistic Heavy Ion Collider (RHIC)~\cite{starridge}, interpreted by some as evidence for a dense, flowing QCD medium~\cite{qgp1,qgp2}. The possible correspondence suggests that a dense medium may also form in \pp collisions at 7 TeV.
While the CMS result is intriguing one can ask whether it does signal novel physics in \pp collisions at LHC energies or whether it is simply an extrapolation of \pp phenomena previously observed at 0.2 TeV.

In the present study we make quantitative comparisons between CMS results at 7 TeV and correlation measurements at 0.2 TeV. The general strategy is to extrapolate measured 0.2 TeV \pp correlations to 7 TeV via the energy dependence established below 0.2 TeV and then determine what other alterations of the extrapolated correlation structure are required to describe the 7 TeV data, both minimum-bias data and with cuts applied.

We address a central question: Is the CMS ``ridge'' a truly novel manifestation  at 7 TeV, suggesting anomalous physics in \pp collisions previously encountered in RHIC heavy ion collisions at 0.2 TeV? Or is the unexpected structure simply explained as an azimuth quadrupole component consistent with RHIC quadrupole systematics extrapolated to 7 TeV and CMS cut conditions?

The paper is arranged as follows. We briefly review the CMS angular correlation results. We then describe general correlation analysis methods applied to nuclear collisions at the RHIC. We review the systematics of angular correlations from 0.2 TeV \pp collisions. We then report quantitative A-B comparisons between 0.2 TeV \pp correlations extrapolated to 7 TeV and the CMS results. We conclude that the CMS ``ridge'' is consistent with RHIC correlation data suitably extrapolated.

 \section{CMS Correlation Analysis}

The CMS analysis is directly related to correlation analysis previously carried out at the RHIC. In order to distinguish what is truly novel at the LHC it is important to establish the quantitative relationships among different analysis methods, and the \pp phenomenology that has emerged from previous work at lower energies.

\subsection{CMS analysis method}

The CMS analysis is based  on normalized pair densities
\bea
S_N &=& \frac{1}{N(N-1)} \frac{d^2 N^{signal}}{d\Delta \eta\, d\Delta \phi} \\ \nonumber
B_N &=& \frac{1}{N^2} \frac{d^2 N^{bkd}}{d\Delta \eta\, d\Delta \phi},
\eea
where for example $N^{signal}$ is a {pair} number, to be distinguished from $N$, the number of (charged) particles in the angular acceptance.
Pair densities $S_N$ and $B_N$ are equivalent to pair numbers $\hat n_{ab}$ defined in Ref.~\cite{axialci,axialcd}, with sibling-to-mixed pair ratio $\hat r_{ab} = \hat n_{ab,sib} / \hat n_{ab,mix}$ for 2D bins $(a,b)$ on difference variables $\eta_\Delta = \eta_1 - \eta_2$ (pseudorapidity) and $\phi_\Delta = \phi_1 - \phi_2$ (azimuth).
The CMS correlation measure is
\bea \label{corrmeas}
R(\Delta \eta, \Delta \phi) &=& \left\langle  \left(\langle N \rangle-1\right) \left(\frac{S_N}{B_N} - 1 \right)   \right\rangle_N \\ \nonumber
&=&  \left(\langle N\rangle -1\right)(\langle \hat r \rangle-1),
\eea
with the correspondence to measure $\langle N \rangle (\langle \hat r \rangle - 1)$ of Refs.~\cite{axialci,axialcd} established in the second line. The correlation measure defined in Eq.~(\ref{corrmeas}) is directly proportional to the specific detector angular acceptance. 

Event multiplicity selection strongly influences spectrum and correlation structure in \pp collisions, mainly by biasing the jet frequency per \pp collision~\cite{ppprd}. In the CMS analysis $p_t$-integral angular correlations were obtained for minimum-bias data and for a cut on ``offline tracks'' $N_{trk} > 110$ ($p_t > 0.4$ GeV/c, $|\eta|< 2.4$) which corresponds to $\langle N_{trk} \rangle = 118$. The corresponding corrected angular density is $dN_{ch}/ d\eta \approx 40$. The NSD density is  $dN_{ch}/ d\eta \approx 5.8$, with $\langle N_{trk} \rangle \approx 16$~\cite{cmsspect}. The ratio of cut-selected to minimum-bias corrected multiplicity is then about 7. $p_t$-differential correlations were also obtained for specific $p_t$ bins. $p_t$ cuts should  modify jet correlations and possibly the nonjet azimuth quadrupole.




\subsection{CMS correlation measurements}

The main results are briefly summarized here. The correlation histograms are discussed in more detail in Sec.~\ref{cmshistos}. The CMS study concluded that jet correlations are enhanced in high-multiplicity collisions. The away-side  ridge (interjet correlations) is flat for $|\eta| < 1.5$, consistent with STAR measurements within the TPC acceptance $|\eta| < 1$~\cite{axialci,daugherity}. A CMS breakdown of \pp angular correlation structure at the LHC can be compared with 200 GeV \pp phenomenology described in Sec.~\ref{200gev}.

Several $p_t$ cut intervals were defined. For $p_t \in [1,3]$ GeV/c and high-multiplicity cut a same-side ``ridge'' is  assumed to be associated with the same-side 2D jet peak.  No corresponding ridge is observed in PYTHIA data. The $p_t$ and $N_{ch}$ dependence of the same-side ridge is studied via ZYAM subtraction, usually applied to jet studies with 1D dihadron correlations~\cite{tzyam}. The same-side ridge does not depend on charge combinations. A similar structure is observed for correlated $\gamma$s from  $\pi^0$ decay. Observation of a same-side ridge in \pp collisions at 7 TeV and possible implications in relation to a same-side ``ridge'' reported in more-central RHIC \auau collisions are the featured results of the CMS analysis.

 \section{Analysis method for this study}

We review technical aspects of STAR correlation analysis applied to nuclear collisions at the RHIC. Method details are provided in Refs.~\cite{porter2,porter3,inverse,axialci,daugherity,davidhq,davidhq2,davidaustin}.


\subsection{Angular correlations on $\bf (\eta_\Delta,\phi_\Delta)$}

Two-particle angular correlations are defined on 4D momentum subspace $(\eta_1,\eta_2,\phi_1,\phi_2)$. In acceptance intervals where correlation structure is invariant on mean position (e.g. $\eta_\Sigma = \eta_1 + \eta_2$) angular correlations can be {\em projected by averaging} onto difference variables (e.g. $\eta_\Delta = \eta_1 - \eta_2$) without loss of information to form {\em angular autocorrelations}~\cite{axialcd,inverse}. The 2D subspace ($\eta_\Delta,\phi_\Delta$) is then visualized. The notation $x_\Delta$ rather than $\Delta x$ for difference variables is adopted to conform to mathematical notation conventions and to reserve $\Delta x$ as a measure of the detector acceptance on parameter $x$. Angular correlations can be formed separately for like-sign (LS) and unlike-sign (US) charge combinations, as well as for the charge-independent (CI = LS + US) combination~\cite{axialci,axialcd}.

\subsection{Correlations on $\bf y_t \times y_t$}

2D correlations on $p_t$ or transverse rapidity $y_t = \ln[(p_t + m_t) / m_\pi]$ ($m_\pi$ for unidentified hadrons) are complementary to 4D angular correlations in 6D two-particle momentum space. $y_t$ is preferred for visualizing correlation structure on transverse momentum. Similar to angular correlations, $y_t \times y_t$ correlations can be defined for LS and US charge combinations but also for same-side (SS, $|\phi_\Delta| < \pi/2$) and away-side (AS,  $ \pi/2 < |\phi_\Delta| < \pi$) subregions of angular correlations. Different correlation mechanisms can be distinguished in the four subspaces~\cite{porter2,porter3}.


\subsection{Correlation measures}


For our initial angular correlation analysis we adopted $\langle  N_{ch} \rangle\, (\langle \hat r \rangle - 1)$ as the correlation measure~\cite{axialci,axialcd}. Sibling and mixed pair numbers are normalized to unit integral, and pair ratio $\hat r$ is averaged over kinematic bins (e.g. multiplicity, $p_t$, vertex position).  That {\em per-particle} measure is an improvement over conventional {\em per-pair} correlation function $C \leftrightarrow \langle r \rangle$ or $\langle r \rangle - 1 \rightarrow \Delta \rho / \rho_{ref}$ ($\Delta \rho$ is the correlated-pair density and $\rho_{ref}$ is the reference- or mixed-pair density). 
The per-particle measure eliminates a trivial $1/ N_{ch}$ trend common to all per-pair measures~\cite{inverse}. However, that {\em extensive} measure is proportional to the specific detector angular acceptance $\Delta \eta \Delta \phi$. 

The corresponding intensive correlation measure is  a form of Pearson's normalized covariance~\cite{inverse},
\bea
\frac{\Delta \rho}{\sqrt{\rho_{ref}}} &=& \frac{\langle N \rangle}{\Delta \eta \Delta \phi}  \, (\langle \hat r \rangle - 1) \rightarrow \rho_0\, (\langle \hat r \rangle - 1)
\eea 
assuming single-particle density $\rho_0$ uniform within the angular acceptance and factorization of the reference density, $\rho_{ref} \approx \rho_0^2$. $\Delta \rho / \sqrt{\rho_{ref}}$  is invariant under combination of uncorrelated parts, therefore should not change with {\em linear superposition} of \nn collisions. The measure is independent of angular acceptance if the underlying physical mechanisms are uniform across the acceptance. 

\subsection{A-B comparisons}

Histograms for this study were binned as $25 \times 28$ on $(\eta_\Delta,\phi_\Delta)$ to match the CMS binning. Detailed comparisons of contour lines between data and model functions then permit quantitative inference of model parameters from the CMS data, typically accurate to 10\% for the simple structures in \pp correlations. The CMS color palette includes 20 colors, whereas this study employs over 50 colors, causing significant differences in shading in some regions. An attempt was made to insure that the intervals spanned by vertical scales are nearly the same for model and CMS data in each comparison, although the offsets may differ due to analysis details. By careful A-B comparisons a good approximation to direct $\chi^2$ model fits to histogram data can be achieved.

 \section{0.2 TeV $\bf p$-$\bf p$ angular correlations} \label{200gev}

\pp and \auau angular correlations at RHIC have been extensively studied by the STAR collaboration~\cite{porter2,porter3,axialci,axialcd,daugherity,davidhq,davidhq2,davidaustin}. Those results provide an essential reference for the CMS \pp measurements at 7 TeV.

\subsection{Two-component angular correlations}

Spectra and correlations in nuclear collisions can be decomposed (near mid-rapidity) into soft and hard components, denoting respectively longitudinal fragmentation (mainly diffractive dissociation) of projectile nucleons and transverse fragmentation of large-angle scattered partons~\cite{ppprd,hardspec,porter2,porter3}. 
The hard-component fraction amounts to a few percent in minimum-bias (NSD) \pp collisions~\cite{ppprd} but increases to about 1/3 of the final-state yield in central \auau collisions~\cite{jetspec}.

\subsection{Minijet phenomenology} \label{phenom}

Minijets play a key role in RHIC collisions~\cite{minijets,hardspec,fragevo,nohydro}. The history of minijets dates from the UA1 observation in 1985 of jet production at $\sqrt{s} = 200$ GeV following pQCD cross-section predictions down to 5 GeV (background corrected to 3-4 GeV)~\cite{ua1}. Because of the parton spectrum structure (power law with lower cutoff near 3 GeV) the minimum-bias parton (jet) spectrum is dominated by 3 GeV jets~\cite{fragevo}. Thus, correlations from 3 GeV minijets and from minimum-bias jets are essentially equivalent.

The interplay of correlations on $(\eta_\Delta,\phi_\Delta)$ and $y_t \times y_t$ for SS/AS angular subregions and LS/US charge combinations distinguishes soft- and hard-component correlation structure~\cite{porter2,porter3}. For instance, SS-US $ y_t \times y_t$ correlations  show clear {\em intra}jet structure extending down to 0.3 GeV/c with mode at  $y_t = 2.7$ ($p_t \sim 1$ GeV/c), corresponding closely (when projected) to the \pp $p_t$ spectrum hard component first revealed in Ref.~\cite{ppprd}. The SS-LS combination is dominated by HBT correlations appearing below 0.5 GeV/c, with negligible jet contribution. 

AS-LS and AS-US correlations contain identical {\em inter}jet contributions appearing above 0.7 GeV/c, also with mode at $y_t = 2.7$ ($p_t \sim 1$ GeV/c). The absence of charge dependence for AS jet structure is consistent with scattered partons (mainly gluons) having no charge correlation. The AS-US combination also shows a soft-component contribution below 0.5 GeV/c.  Cuts on $y_t \times y_t$ in turn isolate corresponding elements of angular correlations. The combination provides a clear picture of local charge conservation in two orthogonal fragmentation processes plus like-sign quantum correlations (HBT)~\cite{porter2,porter3}.


\subsection{Correlation model functions}

The soft component of angular correlations is modeled by a 1D Gaussian on $\eta_\Delta$ with r.m.s.\ width approximately 1. The model is assumed to be uniform on $\phi_\Delta$ for simplicity. However, there are indications that the soft component is suppressed near $\phi_\Delta = 0$ as a result of local transverse-momentum conservation. The soft component falls within $p_t < 0.5$ GeV/c. Its amplitude decreases to zero with increasing centrality in \aa collisions. The soft component is exclusively US pairs, reflecting local charge conservation during projectile-nucleon fragmentation.

The correlation hard component has two parts, a SS 2D peak at the origin and an AS 1D peak on azimuth uniform on $\eta_\Delta$ (within the STAR TPC acceptance). The SS jet peak (intrajet correlations) is well modeled by a 2D Gaussian. The SS peak is also characterized as $\rho_0(b) j^2(\eta_\Delta, \phi_\Delta, b)$ within angular acceptance $(\Delta \eta,\Delta \phi) = (2,2\pi)$---the STAR TPC acceptance~\cite{jetspec}. The corresponding CMS acceptance is $(4.8,2\pi)$. Except for \pp and most-peripheral \aa collisions the AS peak is conveniently modeled as an AS dipole based on $\cos(\phi_\Delta - \pi)$.


The combined model function~\cite{axialci,daugherity}, including the azimuth quadrupole term $\cos(2\phi_\Delta)$ required to describe minimum-bias \aa angular correlations, is
\bea \label{modeleq}
\frac{\Delta \rho}{\sqrt{\rho_{ref}}} \hspace{-.02in}  & = & \hspace{-.02in} A_0 + A_1 \, e^{- \frac{1}{2} \left\{ \left( \frac{\phi_{\Delta}}{ \sigma_{\phi_{\Delta}}} \right)^2 \hspace{-.05in}  + \left( \frac{\eta_{\Delta}}{ \sigma_{\eta_{\Delta}}} \right)^2 \right\}} 
+ A_2\, e^{-\frac{1}{2} \left( \frac{\eta_{\Delta}}{ \sigma_{2}} \right)^2  } \nonumber \\
&+&   
A_{D}\, \left[1+\cos(\phi_\Delta - \pi)\right]/2 + A_{Q}\, 2\cos(2\, \phi_\Delta),
\eea
where a narrow 2D exponential describing quantum correlations and electron pairs from $\gamma$ conversions has been omitted for clarity. The dipole term in Eq.~(\ref{modeleq}) is defined differently from that in Ref.~\cite{daugherity}. 

For \pp collisions the AS dipole term (interjet correlations) may be replaced by a 1D Gaussian on azimuth centered at $\pi$ with periodic image at $-\pi$~\cite{tzyam}.  The total quadrupole inferred from Eq.~(\ref{modeleq}) then includes the {\em nonjet} quadrupole and a possible quadrupole contribution from the AS jet peak. For minimum-bias \aa collisions the latter is small, since for r.m.s.\ widths greater than 1.2 the higher Fourier components of the AS periodic peak array (including the quadrupole term) approach zero~\cite{tzyam}.

The minimum-bias SS 2D jet peak in \pp collisions is strongly elongated in the {\em azimuth} direction, with approximate 3:2 aspect ratio~\cite{porter2,porter3}. The strong $\phi$ elongation in \pp collisions contrasts with strong {$\eta$} elongation in more-central \auau collisions, with 3:1 aspect ratio~\cite{axialci,daugherity}.

\subsection{200 GeV \pp model parameters}

\begin{figure}[h]
  \includegraphics[width=1.65in,height=1.6in]{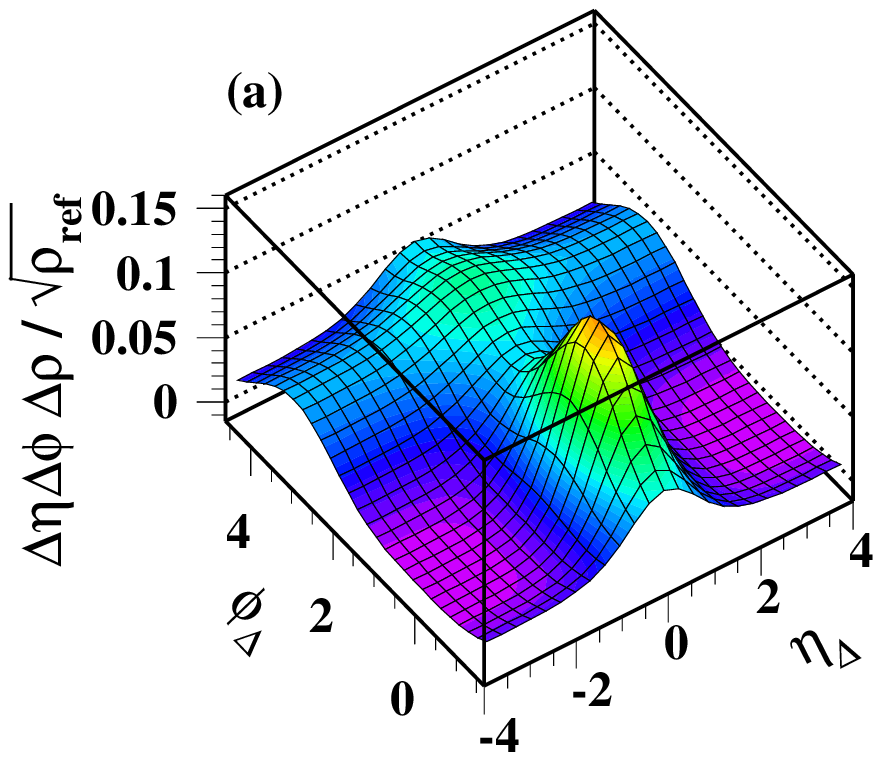}
  \includegraphics[width=1.65in,height=1.6in]{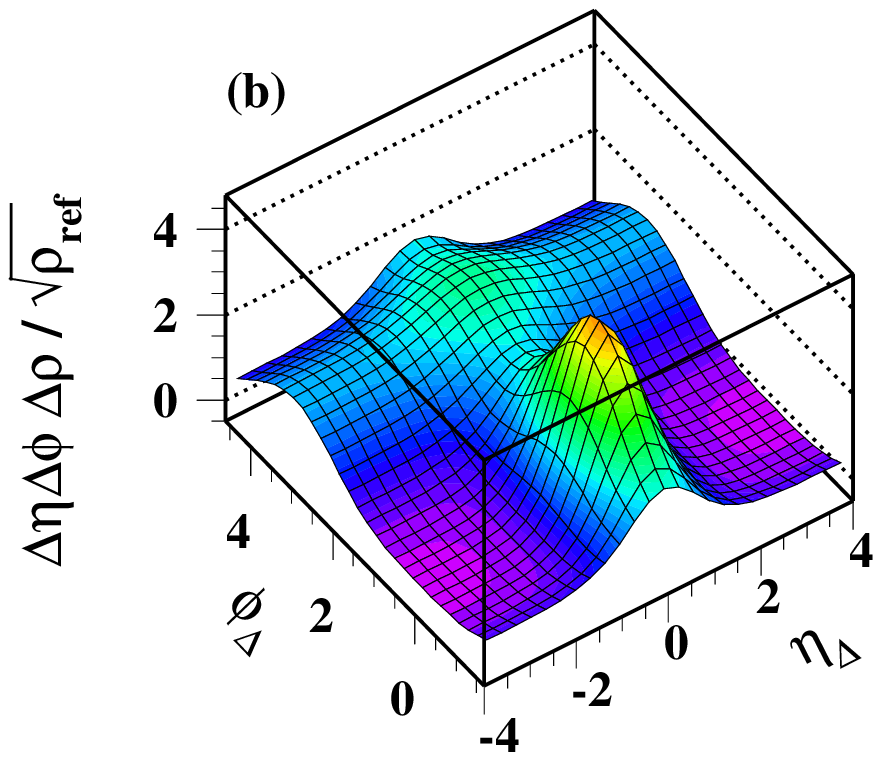}
\caption{\label{fig1} (Color online)
(a) Model angular correlations representing 200 GeV NSD \pp collisions~\cite{porter2,porter3}. 
(b) The same plot rescaled by the CMS angular acceptance $\Delta \eta \Delta \phi = 4.8 \times 2\pi$.
 } 
 \end{figure}

Figure~\ref{fig1} shows model results for 200 GeV NSD \pp collisions for the STAR intensive measure (left panel) and for the CMS extensive measure including the angular acceptance (right panel), the latter as in Ref.~\cite{axialci}.
Results obtained directly from NSD \pp collisions~\cite{porter2,porter3} are consistent with peripheral \aa collisions extrapolated to \nn collisions~\cite{daugherity}, providing a cross check of methods and data consistency. The two cases are thus equivalent.

The measured 200 GeV \nn correlation parameters are summarized in the first row of Table~\ref{table}. The 2D Gaussian parameters~\cite{daugherity} are $A_1 = 0.058$, $\sigma_{\eta_\Delta} = 0.64$ and $\sigma_{\phi_\Delta} = 0.9$.  The amplitude of the 1D Gaussian on $\eta_\Delta$ is $A_2 = 0.023$, with $\sigma_2 \approx 1$.  The AS azimuth dipole amplitude is $A_D \approx 0.046$ (twice the value reported in Ref.~\cite{daugherity} because of the different dipole definition). When modeled as a 1D Gaussian the AS peak amplitude remains 0.046, but with r.m.s.\ width  $\approx 1$ implying a significant  jet-related quadrupole component.  The nonjet quadrupole amplitude extrapolated from \auau centrality dependence is $A_Q \sim 0.0005$. The coefficients of the dipole and nonjet quadrupole {\em sinusoids} are $A_D / 2 = 0.023$ and $2 A_Q = 0.001$ respectively. The sinusoid coefficients are important for the curvature discussion in Sec.~\ref{curves}.

\subsection{Multiplicity dependence of the hard component} \label{starmult}


The $p_t$-spectrum hard component in 200 GeV NSD \pp collisions scales in amplitude (relative to the soft component) nearly linearly with $N_{ch}$ as $N_{ch}$ increases by a factor ten relative to the NSD value~\cite{ppprd}. The spectrum hard component is quantitatively consistent with minijet correlations~\cite{jetspec} and with pQCD-calculated fragment distributions~\cite{fragevo}. The single-particle hard-component yield increases by a factor 50 as the particle multiplicity near mid-rapidity increases ten-fold. Equivalently, the jet frequency per \pp collision within one unit of $\eta$ increases from 2\% to nearly 100\%. The jet frequency saturates at one jet (pair) per collision for large event multiplicities.

Corresponding jet angular correlations scale as follows. The mean jet fragment multiplicity ($\sim 2.5$, dominated by 3 GeV jets) does not change significantly with $N_{ch}$ (the number of correlated pairs per jet is then fixed), but the jet frequency increases by a factor 50 with $10\times$ increase in  $N_{ch}$, rising to one jet per event. Jet correlations measured by $\rho_0\, j^2$ then scale up as $10 \times (50 / 10^2) = 5$, since pair ratio $j^2$ represents correlated pairs / reference pairs. 

\subsection{Collision-energy dependence}

We observe that minijet correlations in the form  $\rho_0(b)\, j^2(\eta_\Delta, \phi_\Delta, b)$~\cite{ptscale,ptedep} and the azimuth quadrupole in the form $\rho_0(b)\, v_2^2(b)$~\cite{gluequad,davidhq} (next subsection) scale with energy approximately as $\log(\sqrt{s_{NN} }/ \sqrt{s_0})$ below 200 GeV, where   $\sqrt{s_0} \approx 13.5$ GeV. We therefore define
\bea \label{rroots}
R(\sqrt{s_{NN}}) \hspace{-.02in} &=& \hspace{-.02in} \log(\sqrt{s_{NN} }/ 13.5~ \text{GeV}) / \log(200/ 13.5)
\eea
to represent collision-energy scaling of jets and quadrupole relative to 200 GeV~\cite{davidhq}. Assuming the same scaling up to 7 TeV implies that both amplitudes should be larger by factor  $\log(7000/ 13.5) / \log(200/ 13.5) \approx 2.3$ at 7 TeV compared to 200 GeV. 

The symbol $R$ introduced in the CMS analysis should be distinguished from energy scaling factor $R(\sqrt{s})$ in Eq.~(\ref{rroots}) defined in Ref.~\cite{davidhq} in connection with $v_2$ analysis.

 \subsection{Azimuth quadrupole systematics}

The azimuth quadrupole, with form $\cos(2 \phi_\Delta)$, is conventionally interpreted in \aa collisions as ``elliptic flow,'' a conjectured response to early development of large pressure gradients in non-central \aa collisions~\cite{hydro}. $v_2$ data inferred from fits to 2D angular correlations and denoted $v_2\{2D\}$, which accurately exclude contributions from jets (``nonflow''), reveal systematic behavior inconsistent with hydro expectations, suggesting an alternative interpretation in terms of interacting gluonic fields~\cite{gluequad,davidhq}. Although the azimuth quadrupole is not usually considered relevant to \pp collisions 
it may explain the CMS ridge manifestation, as demonstrated by this analysis. 


An analysis of $p_t$-integral $v_2\{2D\}$ for \auau collisions at 62 and 200 GeV, combined with SPS $v_2\{EP\}$ data at 17 GeV, led to the following simple relation which describes $v_2\{2D\}$ data from 13.5 to 200 GeV~\cite{davidhq},
\bea \label{v2sys}
\frac{\Delta \rho[2]}{\sqrt{\rho_{ref}}} = \rho_{0}(b)\, v_2^2\{2D\}(b) = 0.0045 \, R(\sqrt{s_{NN}})\,\epsilon^2_{opt}\, n_{bin}.
\eea
For 200 GeV NSD \pp collisions $n_{bin} = 1$ and $\epsilon_{opt} \approx 0.3$, yielding $\rho_{0}\, v_2^2 \approx 0.0005$. For the CMS analysis additional factors (defined above) lead to $2 R_Q \equiv R(\text{7 TeV})\, \Delta \eta \Delta \phi\, 2\rho_0\, v_2^2 = 2.3 \times 30 \times 2 \times 0.0005 = 0.07$  as the {\em predicted} quadrupole amplitude for minimum-bias \pp angular correlations consistent with CMS measure $R$. 


The role of \pp collision centrality is of significant interest at the LHC~\cite{ppcent}.
Requiring increased multiplicity should bias the \pp collision geometry to more-central collisions.  The product $\epsilon^2_{opt}(b)\, n_{bin}(b)$ in Eq.~(\ref{v2sys}) increases rapidly with increasing centrality in \auau collisions to a maximum for mid-central collisions (50-fold increase), beyond which the rapidly decreasing eccentricity dominates~\cite{davidhq}. The centrality trend for \aa collisions is independent of collision energy and absolute system size, suggesting that with increasing multiplicity (centrality) $\rho_{0}\, v_2^2\{2D\}$ for \pp collisions also increases substantially.






\section{7 TeV $\bf p$-$\bf p$ angular correlations} \label{cmshistos}

We compare energy-scaled 0.2 TeV angular correlation results to CMS 7 TeV results for minimum-bias and cut-selected data. The inferred parameters for two energies and several cut conditions are summarized in Table~\ref{table}.

\subsection{Minimum-bias angular correlations}

\begin{figure}[h]
  \includegraphics[width=1.65in,height=1.63in]{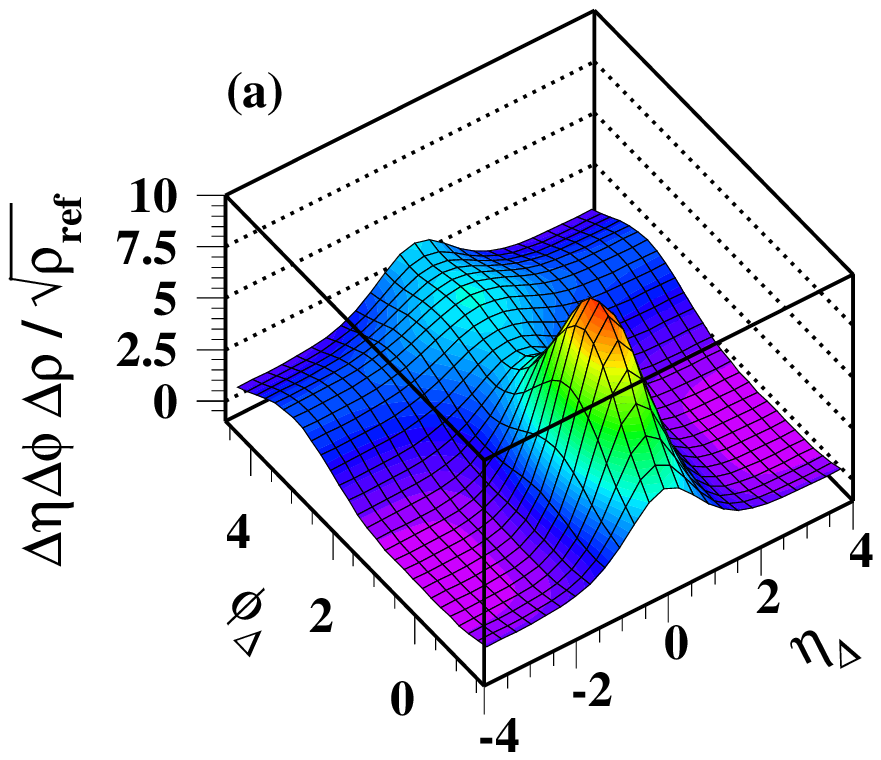}
  \includegraphics[width=1.65in,height=1.65in]{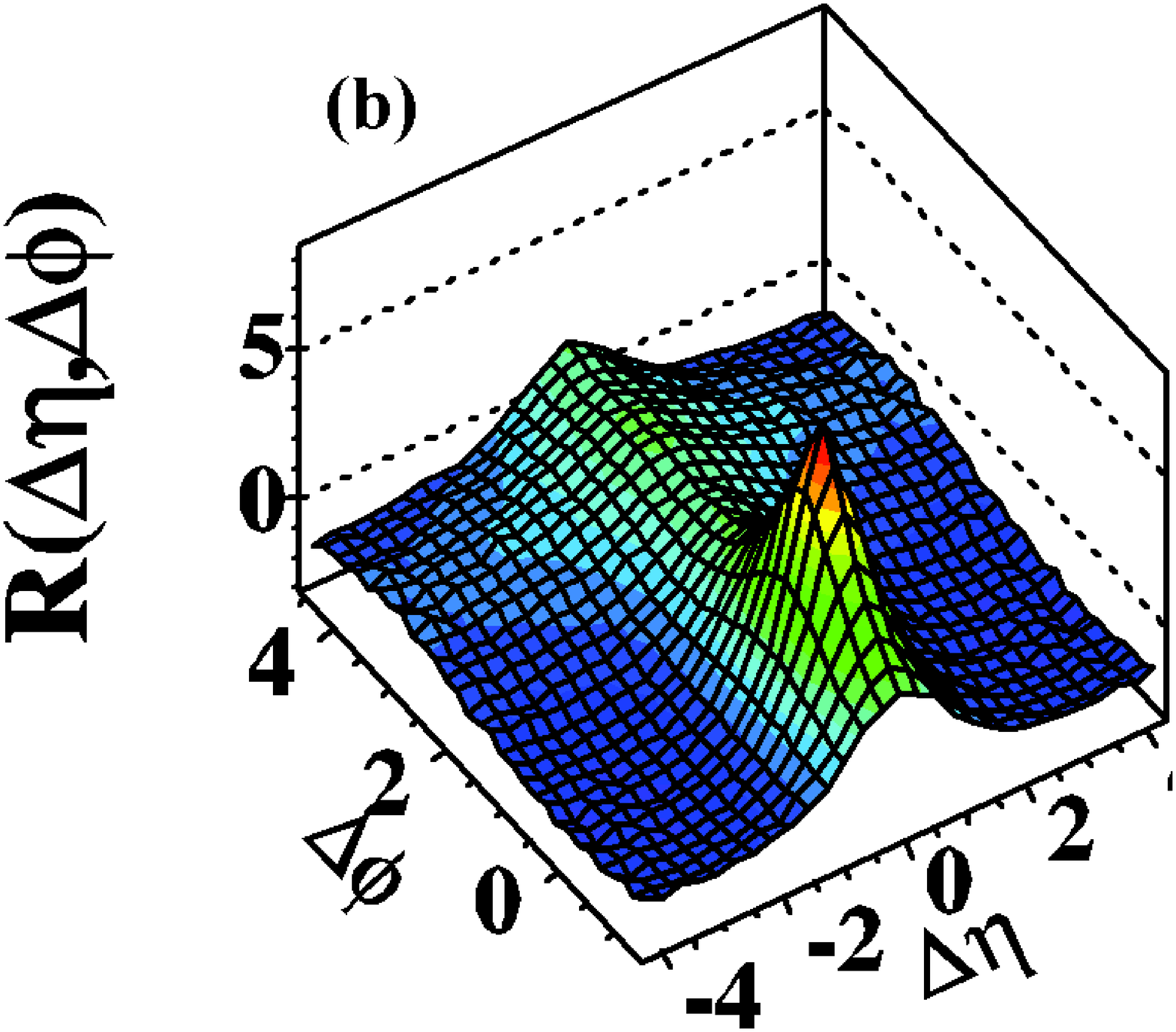}
\caption{\label{fig2} (Color online)
(a) The function from Fig.~\ref{fig1} (left panel) rescaled by energy-dependent factor $R(\text{7 TeV}) = 2.3$.
(b) CMS angular correlations for minimum-bias 7 TeV \pp collisions~\cite{cms}.
 } 
 \end{figure}

For the extrapolation we assume the simplest case: $\log(\sqrt{s})$ scaling observed below 200 GeV continues up to 7 TeV. We then scale all STAR minimum-bias 200 GeV angular correlation structure as $X(\text{7 TeV}) = \left[R(\text{7 TeV})\, 2\pi \Delta \eta\right] \, X(\text{0.2 TeV})$, with $R(\text{7 TeV})$ = 2.3 and $\Delta \eta = 4.8$. The overall scaling factor is therefore $\left[2.3 \times 30\right] \approx 70$. The resulting model in Fig.~\ref{fig2} (left panel) compares well with CMS minimum-bias data in the right panel, but only if the extrapolated AS ridge amplitude is reduced by factor 0.6 (compare with Fig.~\ref{fig1}). The AS amplitude reduction may be a consequence of the larger $4\pi$ $\eta$ acceptance at 7 TeV compared to 200 GeV~\cite{fragevo}. The same reduction factor is retained for all other comparisons.
The large $\phi$ elongation (3:2) of the SS 2D peak observed in 200 GeV \pp collisions~\cite{porter3} persists in 7 TeV collisions, as does the 1D Gaussian on $\eta_\Delta$ associated with longitudinal projectile-nucleon fragmentation~\cite{porter2}.

The actual AS peak model in Fig.~\ref{fig2} (left panel) is a 1D Gaussian with $\sigma_{\phi_\Delta} \approx 1$. The quadrupole component of the Gaussian is then nominally $2A_Q \approx 0.125 \times A_D$~\cite{tzyam}. However, for the purpose of the curvature discussion in Sec.~\ref{curves} we represent the Gaussian as a combination of only AS dipole and quadrupole terms. Requiring an equivalent net curvature at the angular origin $\phi_\Delta = 0$ defines $2A_Q \approx 0.05 A_D$. The jet-related quadrupole then also represents curvature contributions from higher Fourier components, mainly the sextupole. We therefore define jet-related quadrupole $2A_{Q,J} = 0.05 \times A_D$ and nonjet quadrupole  $A_{Q,NJ}$, with total quadrupole $A_Q = A_{Q,NJ} + A_{Q,J}$ inferred from model fits with Eq.~(\ref{modeleq}). Nonjet and jet-related quadrupoles (in that order) are shown in both the $2A_Q$ and the $2 R_Q$ columns of Table~\ref{table}.


\subsection{High-multiplicity cut} \label{highmult}

\begin{figure}[h]
  \includegraphics[width=1.65in,height=1.58in]{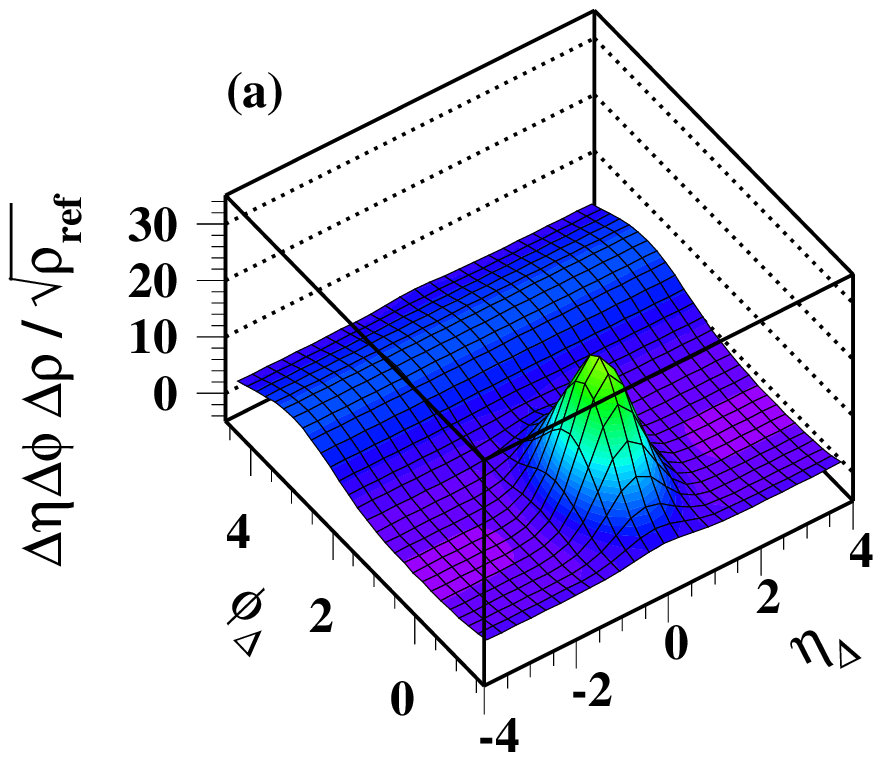}
  \includegraphics[width=1.65in,height=1.65in]{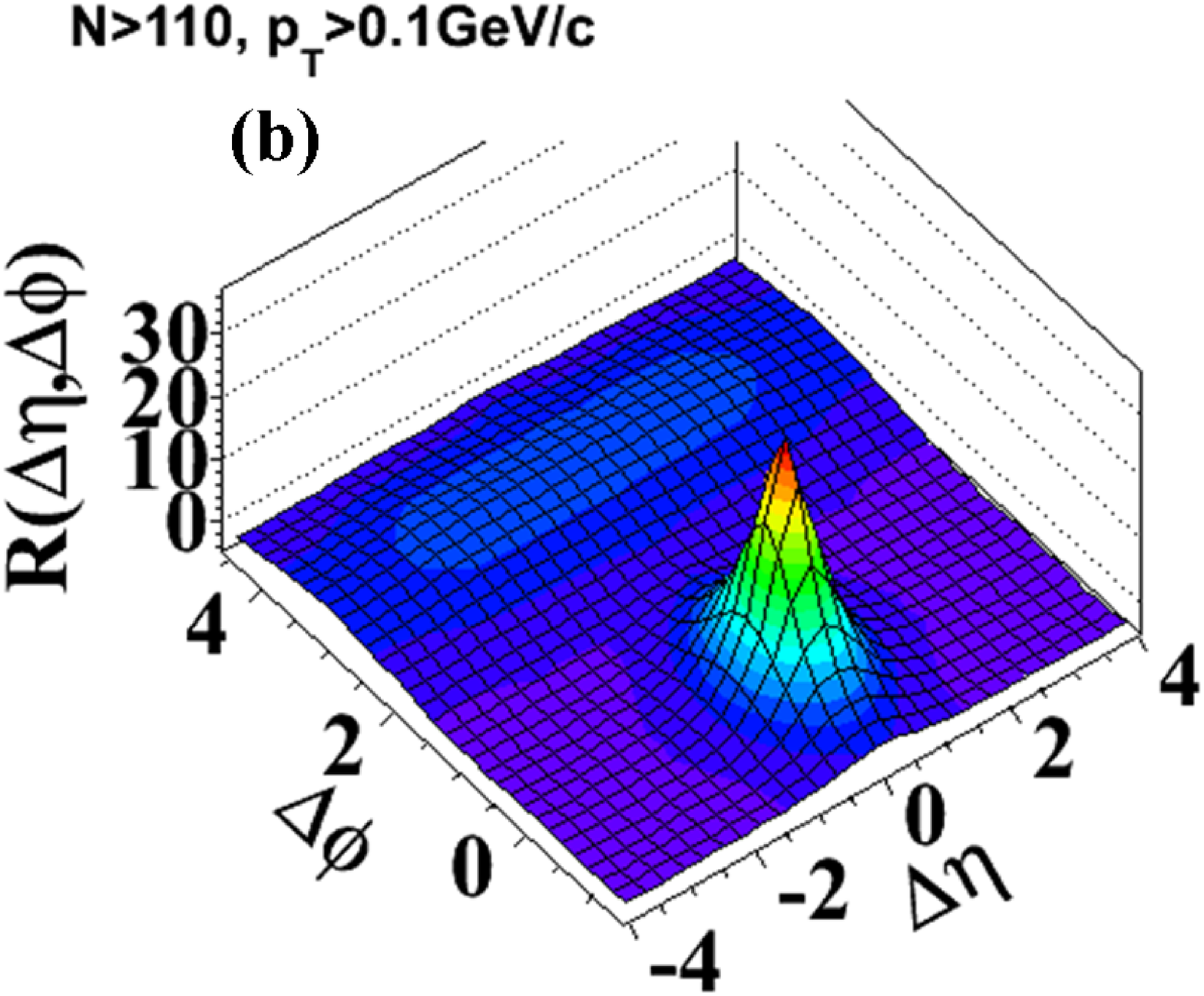}
\caption{\label{fig3} (Color online)
(a) Model function from Fig.~\ref{fig2} (left panel) multiplied by factor 3, with SS peak narrowed on $\eta_\Delta$ and with 1D $\eta_\Delta$ Gaussian removed.
(b) CMS $p_t$-integral angular correlations for high-multiplicity 7 TeV \pp collisions~\cite{cms}.
 } 
 \end{figure}

Selecting high-multiplicity \pp collisions biases the jet frequency per event to larger values~\cite{ppprd}. The increased hard scattering can be interpreted as a result of increased \pp centrality induced by the multiplicity cut.
Figure~\ref{fig3} (left panel) shows the model function in Fig.~\ref{fig2} (left panel) with the following changes: (i) The jet structure (SS 2D peak, AS ridge) is scaled up by factor 3, (ii) the SS peak $\phi$ width is reduced from 0.9 to 0.65 and the $\eta$ width from 0.64 to 0.58, and (iii) the 1D $\eta$ Gaussian is eliminated. The quadrupole component for this figure (not visible in this plotting format) is unchanged from Fig.~\ref{fig2} (minimum-bias $R_Q = 0.07$). The result compares well with CMS data in the right panel, with the exception of the narrow contribution from quantum correlations and electron pairs at the peak not included in the model. 


 \begin{figure}[t]
  \includegraphics[width=3.3in,height=2.95in]{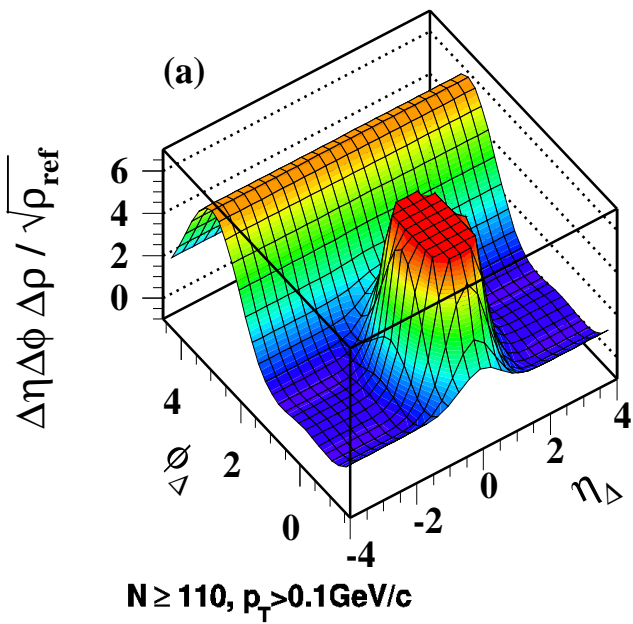}
  \includegraphics[width=3.3in,height=3.in]{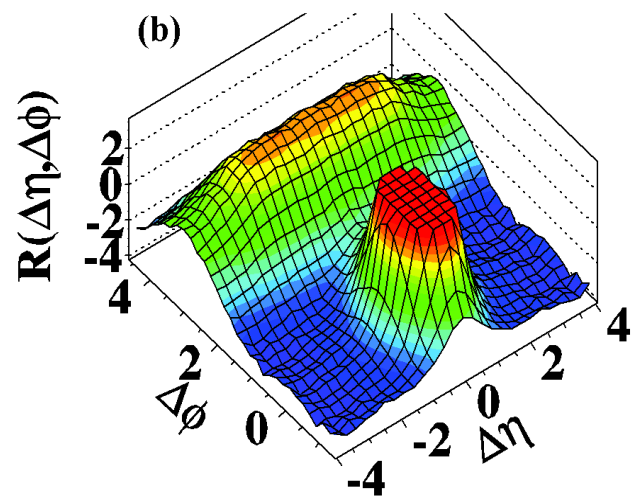}
\caption{\label{fig4} (Color online)
Histograms from Fig.~\ref{fig3} rescaled by factor 5. 
(a) Model function from Fig.~\ref{fig3} (left panel) but with quadrupole amplitude increased by factor 6 to $R_Q = 0.42$. 
(b) CMS $p_t$-integral correlations for high-multiplicity cut~\cite{cms}.
 } 
 \end{figure}

 Figure~\ref{fig4} shows the results in Fig.~\ref{fig3} with the vertical axis range reduced by a factor 5 to enhance small-amplitude details. Correspondence of the SS jet peaks near the base is evident. The AS ridge shows significant reduction in the data at larger $\eta_\Delta$ (lower panel) which is not included in the model function (upper panel) inferred within the STAR TPC acceptance. 

The nonjet quadrupole amplitude is increased by factor 6 for this figure (relative to the minimum-bias value) to $2 R_{Q,NJ} = 0.42$. The larger quadrupole amplitude is required to change the SS curvature in $|\eta_\Delta| > 2$ from concave upward to slightly concave downward (compare with Fig.~\ref{fig3} -- left panel). Thus, the 6-fold quadrupole increase is already required by CMS data prior to imposition of $p_t$ cuts. The curvature on azimuth at $\phi_\Delta = 0$ for $|\eta_\Delta| > 2$ is the sole basis for identification of a ``ridge.'' The curvature issue is further discussed in Sec.~\ref{curves}.



 \begin{figure}[t]
  \includegraphics[width=3.3in,height=2.95in]{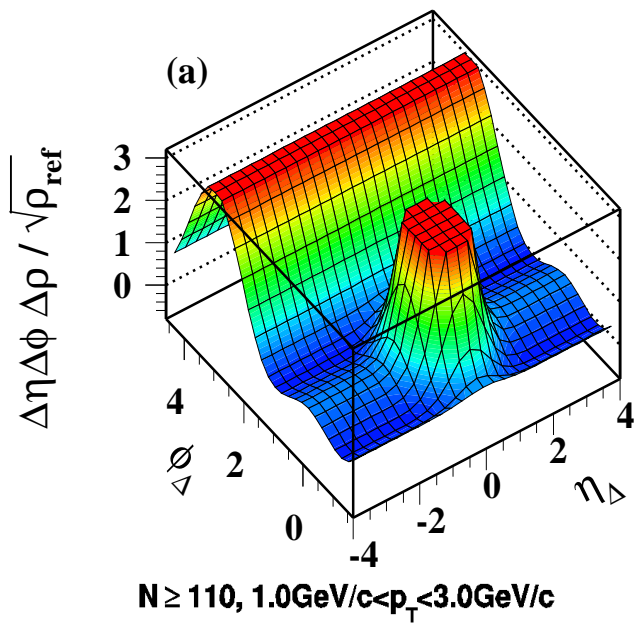}
  \includegraphics[width=3.3in,height=3.in]{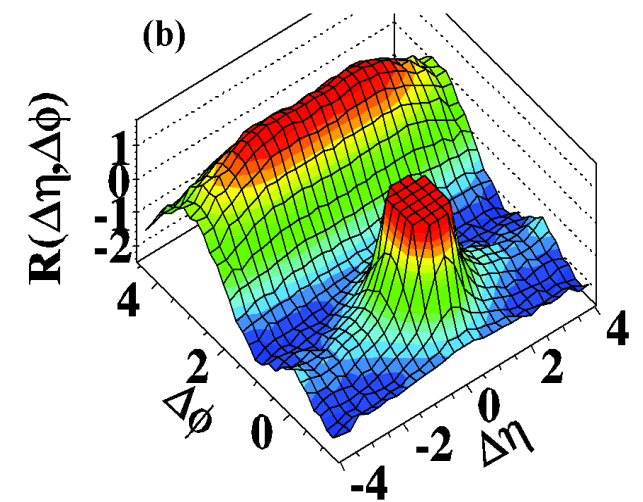}
\caption{\label{fig5} (Color online)
(a) Model function from Fig.~\ref{fig4} (upper panel) with the same quadrupole amplitude but with jet correlation amplitudes and peak widths reduced (see text).
(b) CMS correlations for high-multiplicity cut and $p_t \in [1,3]$ GeV/c~\cite{cms}.
 } 
 \end{figure}

\subsection{High-multiplicity plus high-$\bf p_t$ cuts} \label{highpt}

With high-$p_t$ cuts we expect to eliminate some fraction of the jet contribution, to eliminate any remaining evidence of the soft component and possibly to change the quadrupole amplitude. Figure~\ref{fig5} (upper panel) shows the model in Fig.~\ref{fig4} (upper panel) with the following changes: (i) The $\phi$ width of the SS 2D peak is further reduced from 0.65 to 0.55 leading to a symmetric SS peak, (ii) The SS peak amplitude is reduced by factor 1/3 and AS ridge by factor 1/2 compared to the high-multiplicity cut alone (consequence of reduced $p_t$ acceptance for jet-correlated pairs) and (iii) the nonjet quadrupole amplitude remains the same as for the high-multiplicity cut alone---$2 R_{Q,NJ} =$ 0.42 increased by factor 6 from minimum-bias 0.07. The comparison to CMS data in the lower panel shows good agreement, especially near the base of the SS 2D peak.

The nonjet quadrupole remains the same in Figs.~\ref{fig4} and \ref{fig5}, but the jet structure in Fig.~\ref{fig5} is reduced in amplitude by factor 2-3 relative to the quadrupole by the $p_t$ cuts.  Because the vertical sensitivity in Fig.~\ref{fig5} is doubled, because the nonjet quadrupole remains enhanced by at least factor 6, and because the jet structure (particulary the AS ridge) is strongly reduced in amplitude by $p_t$ cuts the SS ``ridge'' structure becomes visually apparent as a change in sign of the azimuth curvature (Sec. ~\ref{curves}).

\begin{table*}
\caption{ \label{table}
\pp correlation systematics. Entries represent systematics inferred from RHIC data at 0.2 TeV and parameters inferred by modeling CMS angular correlation histograms at 7 TeV from Ref.~\cite{cms}.  The left columns indicate collision energy and cut conditions, including minimum-bias (MB) data. The $A_X$ represent parameters from STAR per-particle analysis with an intensive correlation measure. The $R_X$ represent corresponding CMS data including extensive acceptance factor $2\pi \Delta \eta \approx 30$. The two entries in the $2A_Q$ and $2R_Q$ columns correspond to nonjet and jet-related (AS peak) quadrupoles respectively.
} 
\begin{tabular}{|c|c|c|c|c|c|c|c|c|c|c|c|} \hline
 $\sqrt{s}$ (TeV) & Condition & $2A_{Q}$    & $A_D/2$  & $A_{SS} $  & $16 A_Q / A_D$   & $2R_{Q} $    & $R_D/2$  & $R_{SS}  $     & $\sigma_\eta$  & $\sigma_\phi$ \\ \hline
 0.2  & MB & $0.001 + .0023$ & 0.023 & 0.058 &  0.57 & $0.03 + 0.069 $ & 0.69 & 1.74 & 0.64 & 0.9 \\ \hline 
 7  & MB & $0.0023  + 0.0032$ & 0.032  & 0.133 &  0.69 & $0.07 + 0.096 $  & $0.96$ & 4.0 & 0.64 & 0.9 \\ \hline 
 7  &  $n_{ch}  $ cut & $0.014  + 0.0095$ & 0.095  & 0.39 &  0.99 & $0.42 + 0.29 $  & 2.9  & 12.0  & 0.58 & 0.65\\ \hline 
 7  & $p_t$,  $n_{ch}  $  cuts & $0.014  + 0.0047$ & 0.047  & 0.13 & 1.59  & $0.42 + 0.14 $  & 1.45 & 4.0 & 0.55 & 0.55 \\ \hline 
 \end{tabular}
\end{table*}

 \section{Discussion}

We observe that minimum-bias jet structure at the LHC is remarkably similar to that at RHIC, simply scaled up by a $\log(\sqrt{s})$ energy-dependence factor. Possible novelty arises with application of multiplicity and $p_t$ cuts. Does the unanticipated ``ridge'' structure imply novel physics in LHC \pp collisions, or is it also an extrapolation of phenomena observed previously at lower energies?


\subsection{Summary of inferred model parameters}

Table~\ref{table} presents results from extrapolation of RHIC \pp data to 7 TeV and from modeling CMS data histograms for several cut conditions. Uncertainties are typicaly 10\% except for the 0.2 TeV \pp  nonjet quadrupole estimate which is an upper limit. The first row summarizes results for \pp collisions at 0.2 TeV (jet parameters) or extrapolated from \auau centrality trends (quadrupole parameter). STAR ($A_X$) and CMS ($R_X$) parameters are related by $R_X = 2 \pi \Delta \eta A_X \approx 30 A_X$, with $\Delta \eta = 4.8$.

In the second row the SS peak and quadrupole amplitudes are scaled up with energy by factor $R(\sqrt{s}) = 2.3$ inferred from jet and quadrupole energy systematics below 0.2 TeV. AS peak amplitude $A_D$ increases by factor $0.6 \times 2.3$ to match the structures observed in Fig.~\ref{fig2}. In the third row jet amplitudes with applied multiplicity cuts are further increased by factor 3 and quadrupole by factor 6 to match the structures observed in Fig.~\ref{fig4}.

In the fourth row the results of $p_t$ cuts applied to CMS data corresponding to Fig.~\ref{fig5} are presented. The quadrupole amplitude does not change significantly. The SS 2D peak amplitude $A_{SS}$ is reduced by factor 1/3 and the AS dipole amplitude $A_D$ is reduced by factor 1/2.

\subsection{Particle densities in \pp collisions}

The $R(\sqrt{s})$ trend defined in Eq.~(\ref{rroots}) and inferred from angular correlation energy systematics below 200 GeV~\cite{ptedep,davidhq} also describes NSD particle densities above 200 GeV. Given the NSD value $dN_{ch}/d\eta = 2.5$ at 200 GeV, factor $R(\sqrt{s})$ predicts the value 5.75 at 7 TeV,  consistent with the recent CMS measurement $\sim 5.8$~\cite{cmsspect}.

The CMS high-multiplicity cut produces a 7-fold increase over the NSD angular density ($dN_{ch}/d\eta = $ 40 vs 5.8). In a study of multiplicity dependence of 200 GeV \pp $p_t$ spectra~\cite{ppprd} the multiplicity variation included a 10-fold increase ($dN_{ch}/d\eta = $ 25 vs 2.5). The maximum particle density at 200 GeV is thus 60\% of the maximum at 7 TeV---less than a factor two difference. RHIC and LHC \pp multiplicity trends are directly comparable.
%


\subsection{\pp multiplicity trends}

From the CMS data we conclude that both $A_Q$ (quadrupole) and $A_D$ (jets)  increase with increasing \pp multiplicity. Are such trends reasonable given our knowledge of \pp collisions at 0.2 TeV? We have two sources of information about \pp multiplicity trends at 0.2 TeV: 1) direct observations  of  0.2 TeV\pp systematics and 2) arguments by analogy with \auau collision centrality.

\paragraph{\bf \pp multiplicity systematics at 0.2 TeV}

Selection of high-multiplicity \pp events is expected to bias toward more-central collisions~\cite{ppcent}. Two-component analysis of  $p_t$ spectra reveals that with a ten-fold increase in event multiplicity the jet frequency in 200 GeV \pp collisions increases by a factor 50. The increase saturates when the jet probability nears 100\% (Sec.~\ref{starmult})~\cite{ppprd}. 
Jet correlation measurements at 200 GeV confirm that multiplicity cuts increase the jet frequency per collision. The per-particle jet correlation amplitude increases by up to a factor 5, but the mean fragment multiplicity per jet is not significantly altered (the multiplicity cut does not bias minimum-bias parton fragmentation)~\cite{porter2,porter3}.

In Sec.~\ref{highmult} we conclude from CMS data that application of a high-multiplicity cut increases the jet correlation amplitude by factor 3.  At 7 TeV the minimum-bias jet frequency increases by factor $R(\text{7 TeV}) = 2.3$ to almost 5\%. The maximum frequency increase is then 50/2.3 = 22. With a 7-fold increase in event multiplicity we therefore expect the correlation amplitude to increase by factor $7(22/7^2) \approx 3$, which we observe in the CMS data. 

\paragraph{\bf \auau centrality systematics}

The CMS data (SS curvature) suggest a factor-6 increase in the nonjet quadrupole with high-multiplicity cut. Is that consistent with the increase in jet correlations?  The nonjet azimuth quadrupole in \aa collisions below 200 GeV depends only on collision energy and initial geometry in the form $b/b_0$ (the geometry of intersecting spheres), not on absolute system size. The two trends are factorized, as in Eq.~(\ref{v2sys})~\cite{davidhq}.  We argue by analogy that the increase in \pp centrality which leads to increased jet production should produce a comparable or greater quadrupole increase.

The nonjet quadrupole amplitude is expressed in  terms of \aa centrality parameters by $2A_Q = \rho_0(b) v_2^2 \propto  n_{bin} \epsilon_{opt}^2$. The AS ridge (jets) scales with \aa centrality as $A_D / 2 \propto \rho_0(b) j^2 \propto \nu / (1 + x(\nu - 1)) \approx \nu$. The Glauber parameters are related by $n_{bin} \sim \nu^4$ and $n_{part} \sim \nu^3$.  Thus,  $A_Q / A_D$ should scale with centrality as $\nu^3$. In \pp collisions $\nu$ can be interpreted as a thickness measure or interaction length of two overlapping spheres. A quadrople increase with multiplicity twice the jet increase is consistent with measured  \pp and \auau centrality trends.

\subsection{Jet correlations response to $p_t$ cuts}





Minimum-bias SS peak properties for 7 TeV jet angular correlations are quantitatively similar to those at 0.2 TeV. In \pp correlations on $y_t \times y_t$ at the lower energy~\cite{porter2,porter3} SS jet correlations extend down to $p_t =$ 0.3 GeV/c, with mode at 1 GeV/c . Nearly half the SS-correlated pairs appear below the mode. In contrast, AS correlations are cut off near 0.7 GeV/c due to initial-state $k_t$ effects. Thus, a smaller fraction of AS pairs appears below 1 GeV/c.  In Sec.~\ref{highpt} we observed that with $p_t \in [1,3]$ GeV/c cuts imposed the 7 TeV SS peak amplitude is reduced by factor 1/3 and the AS ridge is reduced by factor 1/2, consistent with 0.2 TeV jet structure on $y_t \times y_t$. With the high-multiplicity cut the SS peak azimuth width decreases from 0.9 to 0.65. With the additional high-$p_t$ cut the azimuth width is further reduced to 0.55 leading to a symmetric SS 2D jet peak. 


\subsection{The azimuth quadrupole in \pp collisions}

The nonjet quadrupole in minimum-bias \pp collisions at 200 GeV is ill-defined because of a fitting ambiguity.  The total $\eta_\Delta$-independent quadrupole in \pp 2D angular correlations within the STAR TPC includes a possible nonjet quadrupole and the second Fourier component of the AS jet peak, 
the latter depending on the actual AS peak width~\cite{tzyam}. If the AS peak is modeled as a dipole (width $\sim \pi / 2$) then the fitted quadrupole component serves as an upper limit on the nonjet quadrupole. 

The nonjet quadrupole in \pp collisions is better determined  by extrapolation from peripheral \auau collisions. 
The nonjet quadrupole in \auau collisions at and below 200 GeV has the simple parametrization shown in Eq.~(\ref{v2sys})~\cite{davidhq}, quite different from minijets which undergo a sharp transition on centrality~\cite{daugherity}.  
%
%
The parametrization includes a $\log(\sqrt{s})$ factor which predicts the quadrupole for minimum-bias \pp collisions at 7 TeV.

Like the CMS ridge, the 200 GeV azimuth quadrupole is insensitive to charge combination, with equal amplitude for like-sign and unlike-sign charge pairs~\cite{axialcd}.  
In minimum-bias \pp collisions at 7 TeV the quadrupole is not visually apparent but may be determined by model fits to 2D histogram data. 
With the imposition of multiplicity and $p_t$ cuts to CMS data the extrapolated quadrupole can become visually apparent as a SS ridge. 
The correspondence between quadrupole properties and the SS ridge in CMS data strongly suggests that the ridge is in fact a manifestation of the azimuth quadrupole. 

Interpretation of the SS ridge at 7 TeV as a manifestation of the azimuth quadrupole does not imply that ``elliptic flow'' plays a role in \pp collisions. Instead, appearance of the quadrupole in elementary collisions is consistent with evidence against a hydrodynamic interpretation in more-central \auau collisions~\cite{hardspec,quadspec,nohydro}. Recently-measured systematic trends~\cite{davidhq,davidhq2,davidaustin} suggest that the nonjet quadrupole is a novel QCD phenomenon~\cite{gluequad}.


\subsection{Azimuth curvatures and ``ridge'' phenomena} \label{curves}

Mathematically, the ``ridge'' observed in CMS \pp data results from a competition between two curvatures on $\phi_\Delta$ within $|\eta_\Delta| > 2$ (which excludes most of the SS 2D peak). A ridge appears when azimuth structure near $\phi_\Delta = 0$ becomes significantly concave downward (defined as negative net curvature). In minimum-bias \pp collisions we observe only positive curvatures in that region. A small relative change in correlation amplitudes may result in the qualitative appearance or disappearance of a ``ridge.''

In the context of the present analysis the dominant structures within $|\eta_\Delta| > 2$ are the AS dipole sinusoid $\cos(\phi_\Delta - \pi)$ and the azimuth quadrupole sinusoid $\cos(2\phi_\Delta)$. 
The curvature of $\cos(2\phi_\Delta)$ is four times the curvature of $\cos(\phi_\Delta - \pi)$ (and with opposite sign) at $\phi_\Delta = 0$. 
The net curvature is then determined by the coefficients of the two sinusoids---$A_D / 2$ for the dipole and $2A_Q$ for the quadrupole. 
Zero net curvature corresponds to $4 \times 2 A_Q = A_D / 2$ or $16 A_Q / A_D = 1$. That ratio is included in Table~\ref{table}. A ridge (negative curvature) is observed if the ratio is significantly greater than one.

As revealed in Table~\ref{table} the combination of $p_t$ and multiplicity cuts increases the nonjet quadrupole amplitude by at least a factor 4 relative to the AS jet peak, changing the curvature sign and producing an apparent SS ridge. 
In effect, the azimuth curvature functions as a comparator, switching states from valley to ridge as one amplitude changes value relative to another. A quantitative change is thereby transformed into a qualitative change, interpreted as the emergence of a novel phenomenon.


\subsection{Is the ``ridge'' at RHIC relevant to \pp at LHC?}

Two aspects of the so-called ``ridge'' phenomenon at RHIC can be distinguished: (i) $\eta$ elongation of a monolithic SS jet peak well described by a single 2D Gaussian~\cite{axialci,daugherity} and (ii) claimed development of a separate ridge-like structure beneath a symmetric 2D jet peak~\cite{starridge}. Item (i), well established for untriggered (no $p_t$ cuts) jet correlations and for {some} combinations of $p_t$ cuts, has been referred to as a ``soft ridge,'' although there is no separate ridge per se. Item (ii) is inferred from other combinations of $p_t$ cuts (``triggered'' analysis).

Elongation of the minimum-bias SS peak (i) undergoes a sharp transition on centrality in RHIC \aa collisions (both \auau and \cucu collisions)~\cite{daugherity}. For more-peripheral and \pp collisions the SS peak is strongly elongated in the $\phi$ direction (3:2)~\cite{porter3}. For more-central collisions the SS peak transitions to strong elongation in the $\eta$ direction (3:1)~\cite{axialci,daugherity}. The transition occurs within a small centrality interval. The mechanism is unknown.

Variation of  SS jet peak structure with $p_t$ cuts is complex, depending on the details of fragmentation for different charge-sign combinations, fragment $p_t$s and hadron species. Such details are currently poorly explored. Assignment of certain aspects of peak structure to a distinct ridge phenomenon for some $p_t$ cuts is questionable.

For the CMS data we observe that the SS 2D peak $\eta$ width at 7 TeV has the same value as that at 0.2 TeV and is reduced slightly with applied cuts. Thus, $\eta$ broadening, item (i) observed at RHIC, is not present. The azimuth width varies with cuts just as in 0.2 TeV \pp collisions.

The appearance of a SS ``ridge'' in CMS data for some cut combinations is associated with item (ii) above, given the apparent similarity. However, the absence of item (i) and consistency with known azimuth quadrupole systematics makes interpretation (ii)  unlikely. There is no indication from present data that the CMS ridge is directly associated with the SS 2D peak. A more definitive resolution may come with systematic study of \pbpb centrality systematics in upcoming LHC heavy ion runs.

\section{Summary}

Direct A-B comparisons between CMS 2D angular correlations from 7 TeV \pp collisions and STAR parametrizations of 0.2 TeV collisions with identical histogram formats provide a quantitative model description of the CMS data. The model fits reveal that there is remarkably little difference between 0.2 TeV and 7 TeV minimum-bias \pp collisions. Midrapidity correlation amplitudes and multiplicity densities increase by factor $R(\sqrt{s} = \text{7 TeV}) = 2.3$ over those measured at 0.2 TeV. The minimum-bias correlation structure is otherwise quite similar within present measurement accuracy. 
Response to multiplicity and $p_t$ cuts is also consistent with trends at 0.2 TeV. The change in jet-like correlation amplitudes and same-side 2D peak widths is consistent with STAR correlation analysis at 0.2 TeV. 

The single novel manifestation at the LHC is the appearance of a same-side ``ridge'' for certain multiplicity and $p_t$ cuts. It is conjectured that the ridge structure in 7 TeV \pp collisions may be similar to that observed in more-central \auau collisions at 0.2 TeV, interpreted by some to indicate formation of a dense QCD medium.

However, extrapolation of azimuth quadrupole systematics measured in \auau collisions at 0.2 TeV to 7 TeV \pp collisions predicts a significant quadrupole amplitude in minimum-bias \pp collisions. Although the exact consequences for the quadrupole of multiplicity and $p_t$ cuts applied in the CMS analysis cannot be predicted quantitatively, a six-fold increase over the minimum-bias quadrupole prediction fully accounts for the same-side ridge structure in 7 GeV \pp collisions. A modest increase in the quadrupole amplitude (enhanced by multiplicity cuts) relative to the away-side ridge (reduced by $p_t$ cuts) can reverse the sign of the curvature near the azimuth origin, resulting in an apparent ridge structure.

We conclude that the azimuth quadrupole apparent in more-central \auau collisions at RHIC is directly visualized in \pp angular correlations at the LHC as a curvature reversal resulting from larger collision energy combined with kinematic cuts.  LHC \pp data offer the possibility to study the systematics of the azimuth quadrupole in elementary collisions. Model fits can be applied as in cited references to increase sensitivity to nonjet quadrupole structure. By such means the azimuth quadrupole may be better understood as a QCD phenomenon.

We greatly appreciate extensive discussions on two-particle correlations in nuclear collisions over a number of years with Lanny Ray, Jeff Porter and Duncan Prindle.
This work was supported in part by the Office of Science of the U.S. DOE under grant DE-FG03-97ER41020.


\end{document}